\documentclass{ws-ijmpcs}
\usepackage{subfigure}
\RequirePackage{lineno}

\begin{document}

%\markboth{Zhijun Liang}
%{Performance and operation experience of the Atlas Semiconductor Tracker}
%\catchline{}{}{}{}{}
%
%%%%%%%%%%%%%%%%%%%%%%%%%%%%%%%%%%%%%%%%%%%%%%%%%%%%%%%%%%%%%%%%%%%%

\title{Performance and operation experience of the Atlas Semiconductor Tracker }

\author{Zhijun Liang\footnote{
}, On Behalf of the ATLAS Collaboration}

\address{University of Oxford
\\
zhijun.liang@cern.ch}

\maketitle

%\begin{history}
%%\received{Day Month Year}
%%\revised{Day Month Year}
%\end{history}
%
\begin{abstract}
We report on the operation and performance of the ATLAS Semi-Conductor Tracker (SCT), which has been functioning for 3 years in the high luminosity, high radiation environment of the Large Hadron Collider at CERN.  We also report on the few improvements of the SCT foreseen for the high energy run of the LHC. We find 99.3\% of the SCT modules are operational, the noise occupancy and hit efficiency exceed the design specifications; the alignment is very close to the ideal to allow on-line track reconstruction and invariant mass determination. We will report on the operation and performance of the detector including an overview of the issues encountered. We observe a significant increase in leakage currents from bulk damage due to non-ionizing radiation and make comparisons with the predictions. 
% Valuable lessons for future silicon strip detector projects will be presented.
%
%
%We report on the operation and performance of the ATLAS Semi-Conductor Tracker (SCT), which has been functioning for 3 years in the high luminosity, high radiation environment of the Large Hadron Collider at CERN.  We’ll also report on the few improvements of the SCT foreseen for the high energy run of the LHC. We find 99.3\% of the SCT modules are operational, the noise occupancy and hit efficiency exceed the design specifications; the alignment is very close to the ideal to allow on-line track reconstruction and invariant mass determination. We will report on the operation and performance of the detector including an overview of the issues encountered. We observe a signiﬁcant increase in leakage currents from bulk damage due to non-ionizing radiation and make comparisons with the predictions. 
%We will also cover the time evolution of the key parameters of the strip tracker, including the evolution of noise and gain, the measurement of the Lorentz angle and the tracking efficiency in the harsh LHC environment. Valuable lessons for future silicon strip detector projects will be presented.
%
%\keywords{}
\end{abstract}
%
%\ccode{PACS numbers:}
%
\section{ Introduction}	

The ATLAS detector~\cite{ATLASdet} is a multi-purpose apparatus designed to study a wide range of physics processes at the Large Hadron Collider (LHC) at CERN.
The recent observation of a Higgs-like particle reported by the ATLAS and CMS collaborations is a milestone in particle physics history~\cite{ATLAShiggs,CMShiggs}. The study of the Higgs-like particle rely heavily on the excellent performance of the ATLAS inner detector tracking system. The semiconductor tracker (SCT) is a precision silicon microstrip detector which forms an integral part of this tracking system. The SCT is constructed of 4088 silicon detector modules~\cite{barrel,endcap}, for a total of 6.3 million strips. Each module operates as a stand-alone unit, mechanically, electrically, optically and thermally. The modules are mounted into two types of structures: one barrel, made of 4 cylinders, and two end-cap systems made of 9 disks each. The SCT silicon micro-strip sensors~\cite{sctdet} are processed in the planar p-in-n technology. The signals are processed in the front-end ABCD3T ASICs~\cite{sctABCD}, which use a binary readout architecture. Data is transferred to the off-detector readout electronics via optical fibres.

%The endcaps and barrels comprise 6944 and 8448 sensors respectively. 367 of the barrel sensors are $<100>$, but all other sensors are $<111>$. 74.9\% were manufactured by Hamamatsu, 17.1\% by CIS and the remaining 8\% were CIS oxygen enriched sensors (for the inner radius of the disks).

%The endcaps and barrels comprise 6944 and 8448 sensors respectively. 367 of the barrel sensors are $<100>$, but all other sensors are $<111>$. 74.9\% were manufactured by Hamamatsu, 17.1\% by CIS and the remaining 8\% were CIS oxygen enriched sensors (for the inner radius of the disks). Each module is composed of two pairs of single-sided p-in-n silicon strip sensors, approx- imately 280μm thick. The two sides of a module are glued back-to-back with one another, with a 40 mrad stereo angle between the strip directions on each side, so as to provide a two-dimensional position measurement.

\section{Detector Operations}
The SCT was installed into ATLAS, and was ready for the first LHC proton-proton collisions at a centre of mass energy of 7GeV in March 2010. Since then and until the end of running in February 2013, more than 99\% of the 6.3 million readout channels were operational. During operation several enhancements were introduced into the SCT Data Acquisition System (DAQ)~\cite{sctdaq} in order to avoid potential sources of inefficiency. SCT introduced online monitoring of chip errors in the data and the automatic reconfiguration of the modules with errors. In addition, an automatic global reconfiguration of all SCT module chips every 30 minutes was implemented. There are 90 RODs in the SCT DAQ each of which processes the data for up to 48 modules, if a ROD experiences errors it will exert a BUSY signal which prevents all ATLAS sub-systems from recording data. From 2011 RODs exerting a busy were automatically removed, reconfigured, and reinserted into a run thus having a minimal impact on data taking.

% Thirteen of the defective end-cap modules are due to a cooling pipe failure.
%%
\section{Hit Efficiency}
A high intrinsic hit efficiency is crucial for the operation of the SCT. The intrinsic efficiencies of the SCT modules are measured by extrapolating well-reconstructed tracks through the detector and counting the number of hits on the track and `holes` where a hit would be expected but it is not found. Dead channels were removed from the analysis. The efficiency is higher than 99\% and the fraction of disabled strips in each layer is less than 0.6\% as shown in Fig. 1(a). 

To maintain the high hit efficiency and reduce fake hits, dedicated readout timing calibrations were performed during the collisions runs.  The in-time hit efficiency as a function of readout delay time has been studied for each SCT module as shown in Fig.~1(b).  The readout timing of 4088 modules were optimized and synchronized to 2ns precision level.

\begin{figure}[pb]
\label{hiteff}
  \centering
  \subfigure[]{\includegraphics[width=0.53\columnwidth]{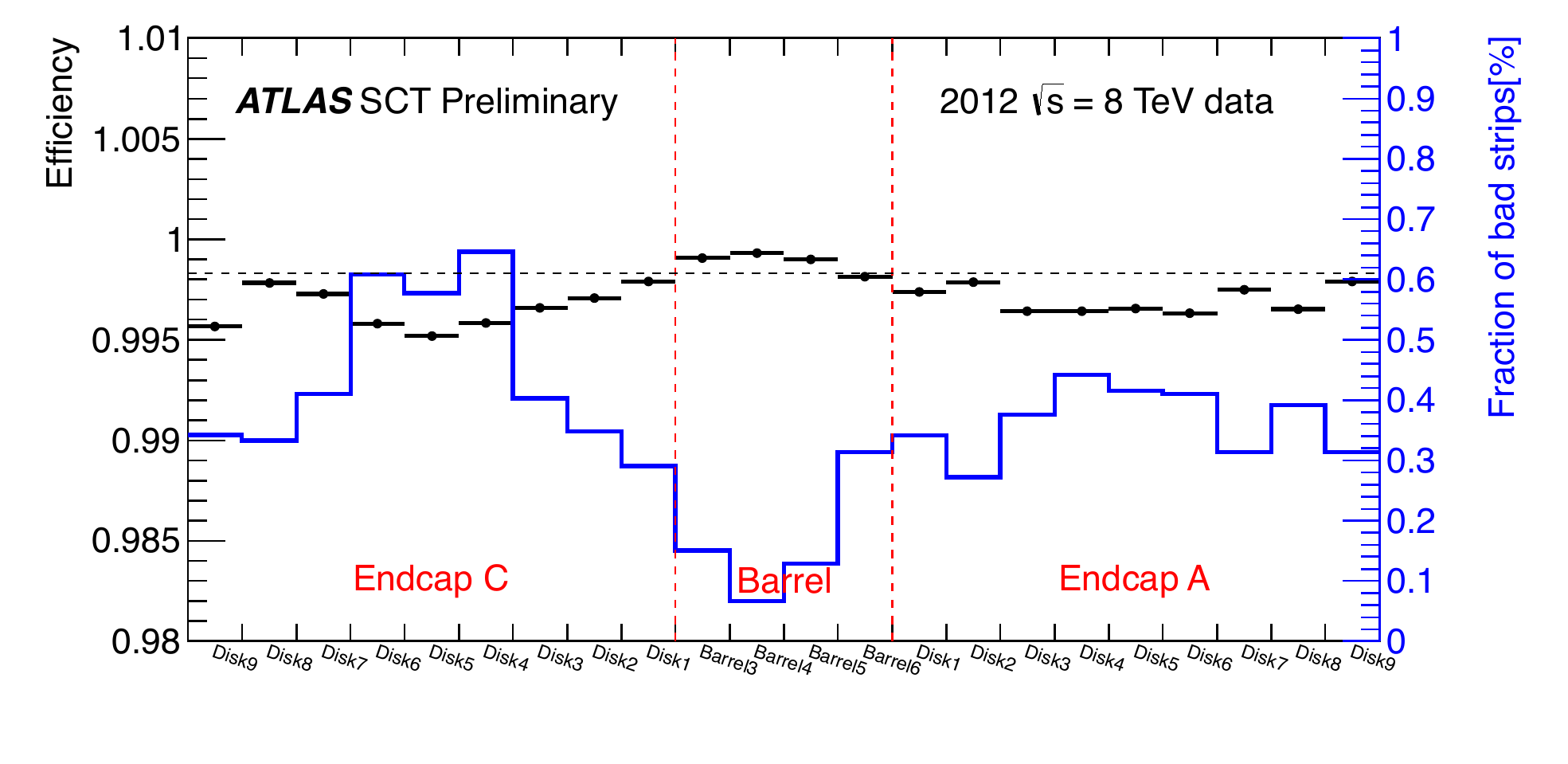}}
  \subfigure[]{\includegraphics[width=0.44\columnwidth]{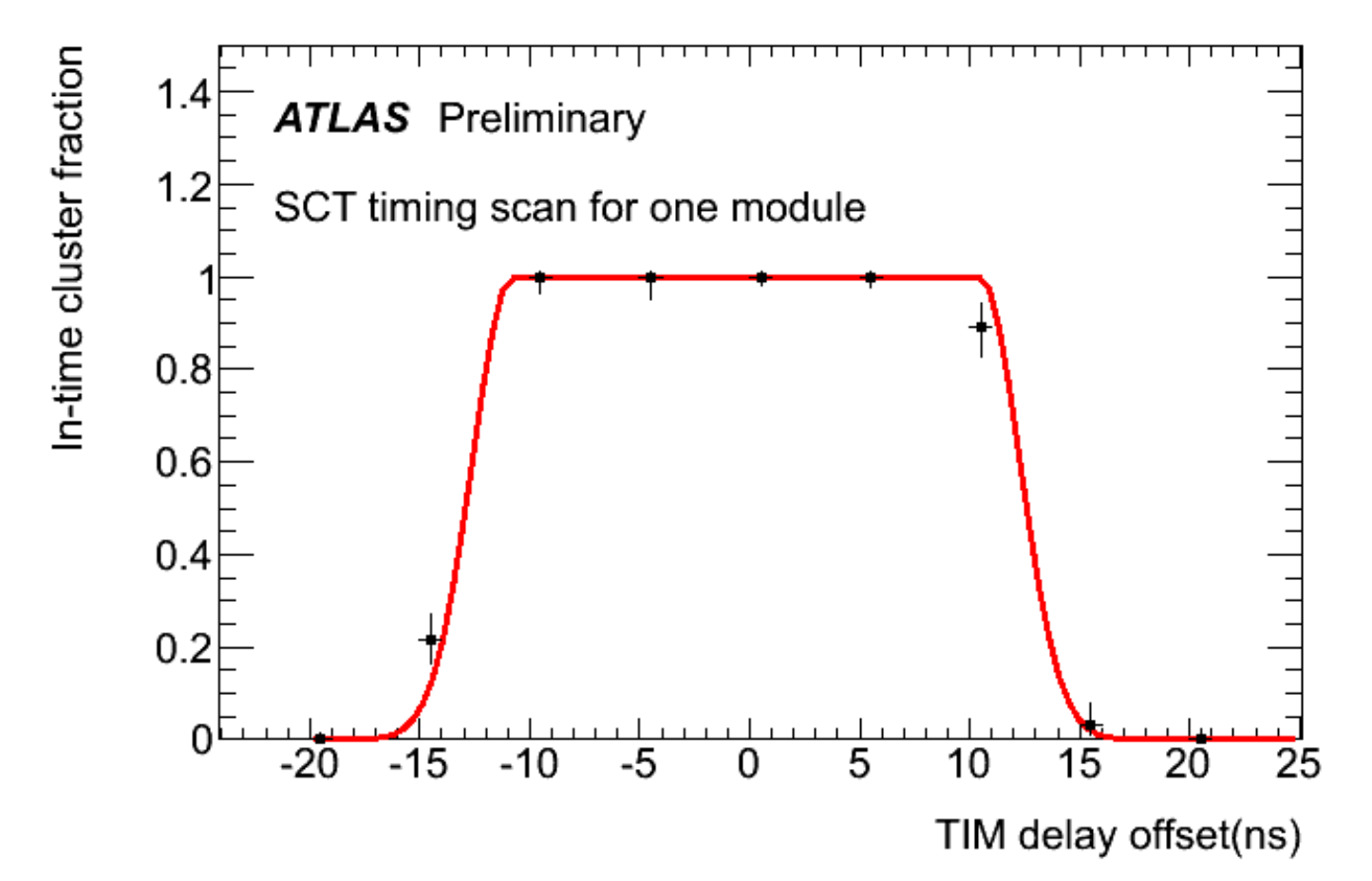}}
\vspace*{8pt}
\caption{ (a) The SCT hit tracking efficiency (black). The blue line and right-hand axis indicates the fraction of disabled strips in each layer. (b) The in-time hit efficiency as a function of readout delay time for one SCT module recorded during a SCT timing calibrations in collisions run.}

\end{figure}

%\begin{figure}[pb]
%\centerline{\includegraphics[width=6.5cm]{timingcurve.pdf}
%%\includegraphics[width=7.7cm]{timingcurve.pdf}
%}
%
%\vspace*{8pt}
%\caption{The SCT hit tracking efficiency (black). The blue line and right-hand axis indicates the fraction of disabled strips in each layer. This figure and all the following are taken from }
%\end{figure}
%
\section{Radiation Damage}
Radiation damage was a key consideration during the design phase of the SCT. Irradiation of silicon sensors results in damage in the bulk silicon and the dielectric layers, with main effects so far being the increase in leakage current of the sensor. Therefore the leakage current has been careful measured during the operation. Fig.~2 shows these measurements as a function of time, along with the integrated luminosity delivered by the LHC on the same time scale. The measured values of the fluence and leakage current are in agreement with predictions from a FLUKA based simulation~\cite{flu}.

%\section{Noise}
%
%\begin{figure}[pb]
%\centerline{
%\includegraphics[width=6.5cm]{noise.png}
%\includegraphics[width=6.5cm]{noise2.png}
%}
%\vspace*{8pt}
%\caption{Luminosity (top), noise (middle) and calibration gains of front-end amplifiers (bottom) as a function of time. Modules are divided into ten different groups according to module type, crystal orientation <111> vs <100> and sensor manufacturer. The green shading and label ‘HI’ indicate periods of heavy-ion running, while extended periods with no beam in the LHC during which the SCT was off are shaded yellow and labelled ‘OFF’. Distributions of chip-averaged noise from response curve tests as of October 2010 (top) and December 2012 (bottom). Noise for eight different types of modules are shown separately.}
%\end{figure}
%

\begin{figure}[pb]
\centerline{
\includegraphics[width=5.5cm]{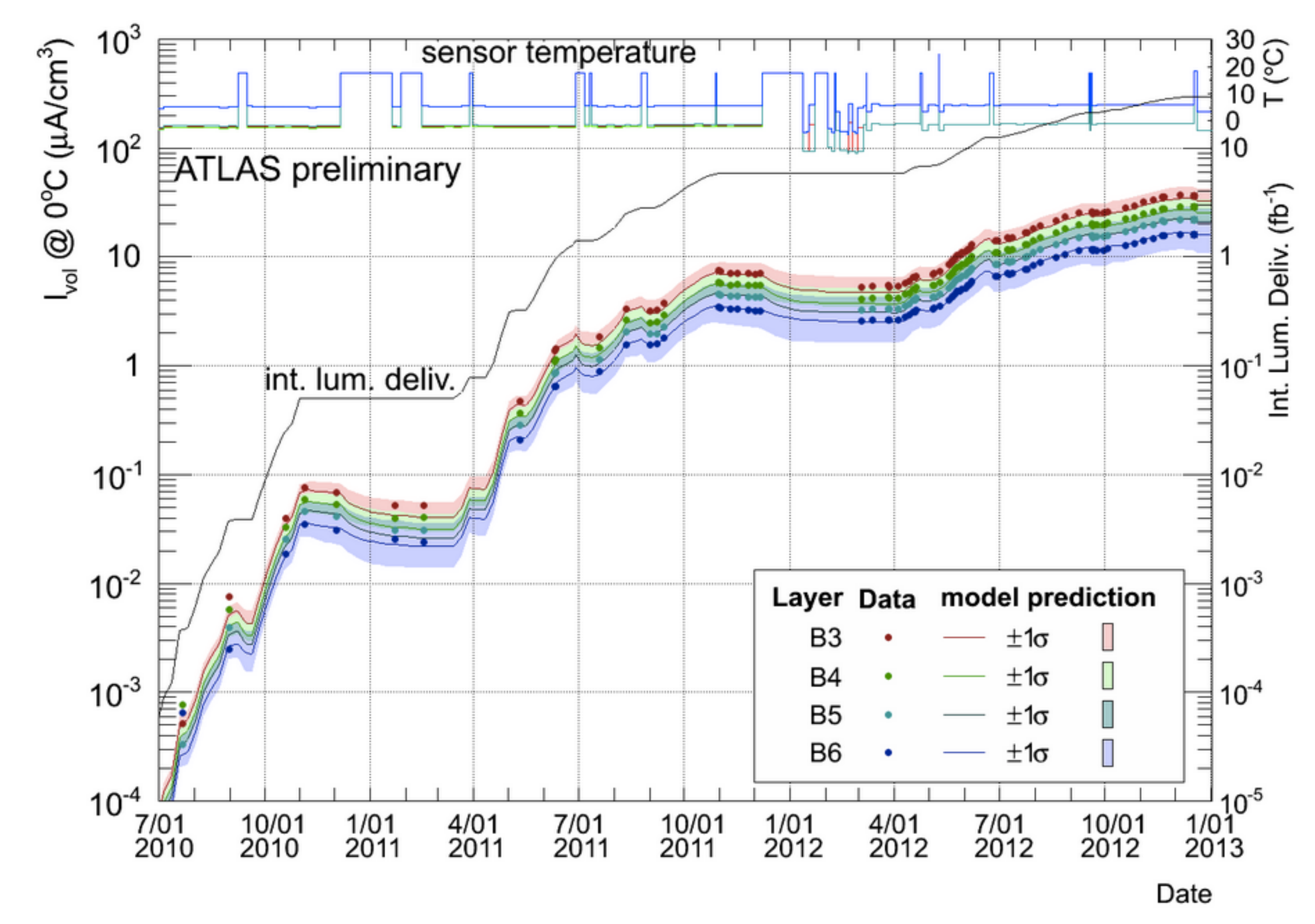}
}
\vspace*{8pt}
\caption{SCT barrel leakage currents during 2010, 2011 and 2012, showing correlations with delivered luminosity and temperature, and compared to predictions from Monte Carlo}
\end{figure}

%\begin{figure}[pb]
%\centerline{
%\includegraphics[width=6.5cm]{lorentz.pdf}
%}
%\vspace*{8pt}
%\caption{}
%\end{figure}
%
\begin{figure}[pb]
\centerline{
  \subfigure[]{\includegraphics[width=0.53\columnwidth]{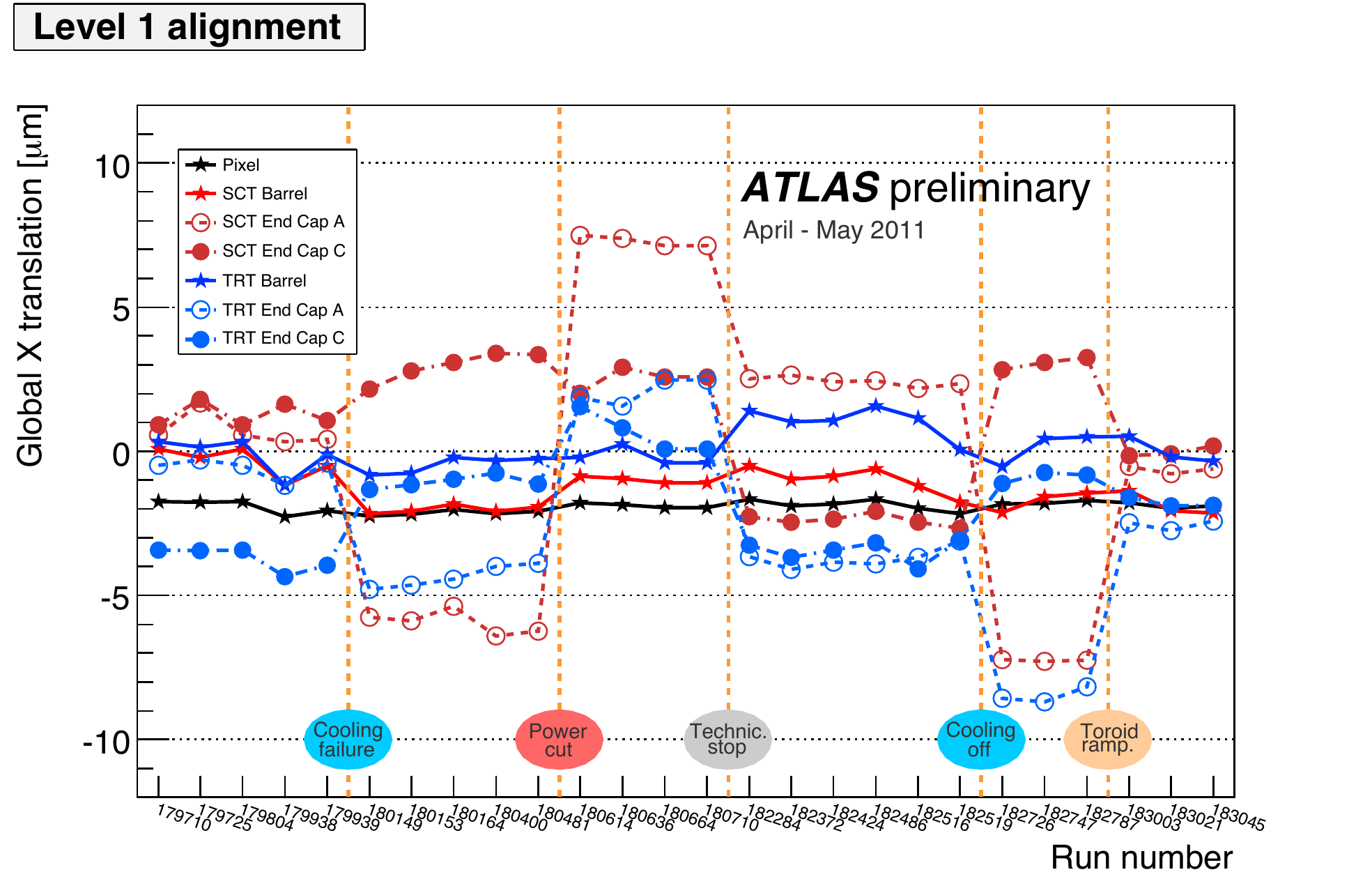}}
  \subfigure[]{\includegraphics[width=0.44\columnwidth]{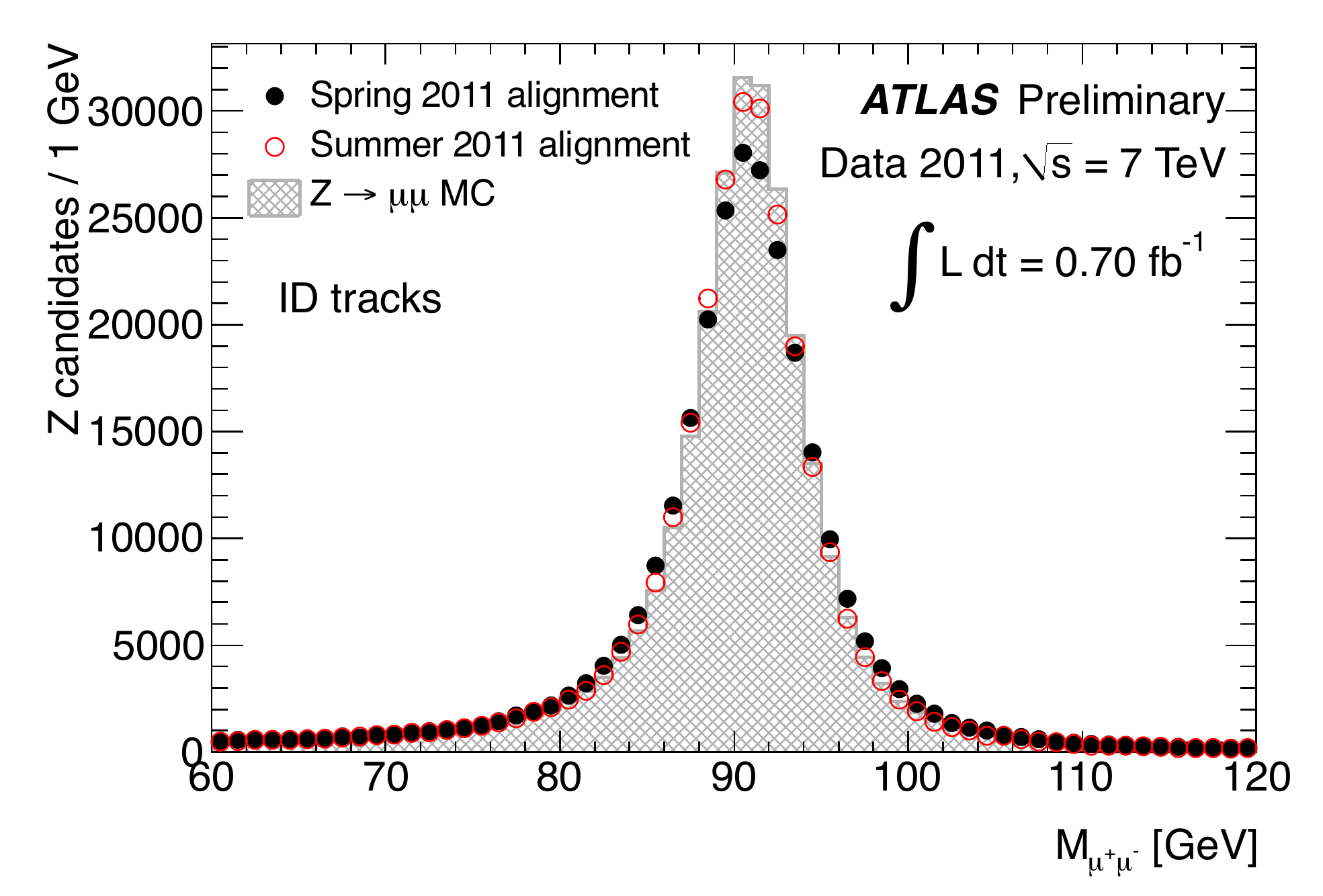}}
}
\vspace*{8pt}
\caption{ (a) Subsystem level, “Level One”, alignment corrections performed on a run by run basis starting from a common set alignment constants. The corrections shown are for translations in the global x direction in different data taking runs. (b) Invariant mass distribution of $Z \to \mu\mu$ decays, where the mass is reconstructed using track parameters from the Inner Detector track of the combined muons only, using about 702 $pb^{-1}$ of data collected during spring 2011. Ideal alignment performance based on Monte Carlo is compared to observed performance of data processed with spring 2011 alignment and data processed with updated alignment constants.
%. In between runs:
%179939 and 180149 there was a cooling system failure
%180481 and 180614 there was a power cut
%180710 and 182284 there was a LHC technical stop where the cooling and power was turned off
%182519 and 182726 a fire alarm went off and as a result the cooling system was required to be shut down
%182787 and 183003 the toroidal magnetic field was dumped
}
\end{figure}

\section{Alignment}
\subsection{Track based Alignment}
Alignment is performed using an algorithm which minimises the the $\chi^2$ between the the measured hit position and the expected position based on on track extrapolation. The alignment is performed at 3 different levels of granularity corresponding to the mechanical layout of the detector: Level 1 corresponds to entire sub-detector barrel and end-caps of Pixel Detector, SCT and TRT. Level 2 deals with silicon barrels and discs, TRT barrel modules and wheels. Level 3 aligns each silicon module and TRT straws having ∼700,000 degrees of freedom in total. Level 1 alignment corrections performed on a run by run basis starting from a common set alignment constants.  As shown in Fig.~3(a) large movements of the detector are measured from track based alignment after hardware incidents. In between these periods, little ($<1\mu$m) movement is observed indicating that the detector is generally very stable. 

After performing three levels of alignment, excellent agreement was found in the residual distributions for both barrel and end cap. The resolution in the $Z \to \mu\mu$ invariant mass distribution from reconstructed tracks, which is shown in Fig.~3(b), is close to the Monte-Carlo expectation.

\subsection{Laser Alignment}

The alignment stability is monitored using a Frequency Scanning Interferometry (FSI) system, which uses data from interferometers attached to the SCT structure to measure changes in length with submicron precision. Fig.~4 shows the effect observed in the FSI during ramping of the solenoid magnetic field in
 December 2009.  Displacements up to 3 $\mu$m are observed while the field is ramping but the detector is observed to return to its initial position at the end of the ramping process.

One of the most important result from FSI measurements is that the SCT detector is found to be stable at
 the $\mu$m level over extended periods of time, in agreement with the track-based alignment results.

\begin{figure}[pb]
\centerline{
\includegraphics[width=7cm]{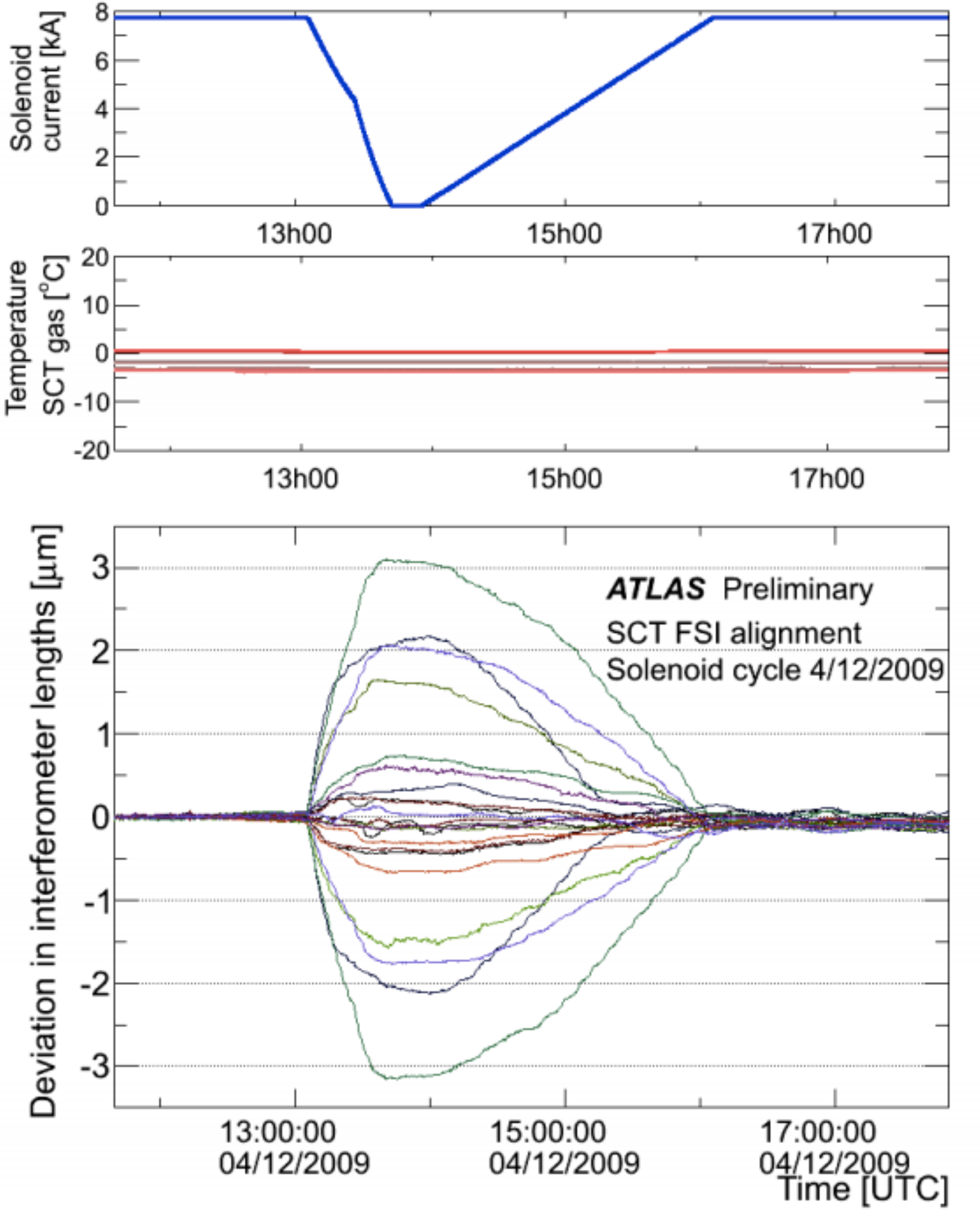}
}
\vspace*{8pt}
\caption{The effect was observed in the FSI during ramping of the solenoid magnetic field in December 2009. The upper plots show the current in the magnets and the temperature of the gas in the SCT; the lower plot shows the detector movements monitored by different laser lines. }
\end{figure}

\section{Summary}
After three years of operation, the SCT is performing well within its design specification in highly challenging conditions. More than 99\% of the detector is still operational. Radiation damage has been observed and is well described by simulation. The detector alignment is very close to the ideal to allow on-line track reconstruction and invariant mass determination.

\end{document}